\documentstyle[preprint,aps]{revtex}
\begin{document}
\title{Microwave-induced thermal escape in Josephson junctions}
\author{N. Gr{\o}nbech-Jensen$^1$, M. G. Castellano$^2$, F. Chiarello$^2$, G.
Torrioli$^2$, M. Cirillo$^3$, L. Filippenko$^{3,4}$, R. Russo$^3$, and C.
Cosmelli$^5$}
\address{$^1$Department of Applied Science, University of California, Davis,\\
California 95616}
\address{$^2$IFN-CNR, via Cineto Romano 42, I-00156 Rome, Italy}
\address{$^3$Department of Physics and INFM, University of Rome "Tor Vergata",\\
I-00133 Rome, Italy}
\address{$^4$Institute of Radio Engineering and Electronics, Mokhovaya, \\
103907 Moscow, Russia}
\address{$^5$Department of Physics and INFN, University of Rome, "La Sapienza",\\
I-00185 Rome, Italy}
\date{\today }
\maketitle

\begin{abstract}
We investigate, by experiments and numerical simulations, thermal activation
processes of Josephson tunnel junctions in the presence of microwave
radiation. When the applied signal resonates with the Josephson plasma
frequency oscillations, the switching current may become multi-valued in a
temperature range far exceeding the classical to quantum crossover
temperature. Plots of the switching currents traced as a function of the
applied signal frequency show very good agreement with the functional forms
expected from Josephson plasma frequency dependencies on the bias current.
Throughout, numerical simulations of the corresponding thermally driven
classical Josephson junction model show very good agreement with the
experimental data.
\end{abstract}


\vskip2pc 

The Josephson tunnel junction is an intriguing solid state physics system
due to the macroscopic quantum nature of the variables describing the
governing equations \cite{Anderson}; irradiation of junctions with microwave
(ac) radiation has produced a number of significant nonlinear phenomena such
as chaos and phase-locking observed both in experiments and theoretical
models. These phenomena have been investigated for almost four decades \cite
{Barone82,Vanduzer98} and successful applications have significantly
benefited from complete theoretical understanding of the relevant dynamics.

Measurements of the escape statistics from the zero-voltage state in
Josephson junctions have been conducted extensively over the years,
successfully confirming consistency with the classic Kramers model for
thermally activated escape from a potential well \cite
{Kramers40,Kramers_stuff}. Escape measurements represent a powerful tool for
probing the nature of potentials and they have also been employed to
discriminate between classical and quantum behavior of the junctions. Work
performed in the past two decades \cite
{Martinis1,Martinis2,Silvestrini97,Ruggiero99} has contributed to generating
general interest toward Josephson junction systems and represented the
background for recent work in the field of quantum coherence and quantum
computing\cite{Martinis3,Han02,Berkley03}. Several groups have
experimentally investigated the effects of applying ac signals to the
junction during measurements of escape out of the Josephson washboard
potential well \cite{Martinis1,Martinis2,Ustinov03}. Applying an ac field to
a low-temperature system has been reported to produce anomalous switching
distributions with signatures of two, or more, distinct dc bias currents for
which switching is likely. These measurements have been interpreted as a
signature of the ac field aiding the population of multiple quantum levels
in a junction, thereby leading to enhancement of the switching probability
for bias currents for which the corresponding quantum levels match the
energy of the microwave photons.

In this letter we report on a systematic investigation of the escape
properties of a Josephson tunnel junction subject to external microwave
radiation and interpret the results classically. We have carried out an
extensive experimental investigation varying the temperature of the samples
from $0.36K$ up to $1.6K$. The results are compared with numerical
simulations of the well-known {\it classical} RSCJ model \cite{Barone82} for
a Josephson junction with appropriate perturbations terms reflecting the
presence of damping, thermal noise, dc and ac bias currents.

Figure 1 illustrates the process under investigation: in the classical
one-degree-of-freedom single-particle washboard potential of the Josephson
junction \cite{Martinis1}, thermal excitations (shaded in the sketch) of
energy $k_BT$ and the energy $E_{ac}$ of forced oscillations due to
microwave radiation, can cause the particle to escape from the potential
well. This process can be traced by sweeping the current-voltage
characteristics of the Josephson junction periodically. Escape from the
potential well corresponds to an abrupt transition from the top of the
Josephson-current zero-voltage state to a non-zero voltage state. The
statistics of the switching events, in the absence of time-varying
perturbations, have been shown to be consistent with Kramers' model \cite
{Kramers40} for thermal escape from a one-dimensional potential. Since the
thermal equilibrium Kramers model does not include the effect of
non-equilibrium force terms, the results of the switching events generated
by the presence of a microwave radiation on a Josephson junction can be
investigated, in a thermal regime, only by a direct numerical simulation of
the governing equation,

\begin{equation}
\frac{\hbar C}{2e}\frac{d^2\varphi }{dt^2}+\frac \hbar {2eR}\frac{d\varphi }{%
dt}+I_c\sin \varphi =I_{dc}+I_{ac}\sin \omega _dt+N(t) \; .  \label{eq:Eq_1}
\end{equation}
Here, $\varphi $ is the phase difference of the quantum mechanical wave
functions of the superconductors defining the Josephson junction, $C$ is the
magnitude of junction capacitance, $R$ is the model shunting resistance, and 
$I_c$ is the critical current, while $I_{dc}$ and $I_{ac}\sin \omega _dt$
represent, respectively, the continuous and alternating bias current flowing
through the junction. The microwave energy sketched in Figure 1 will then be 
$\propto I_{ac}^2$. The term $N(t)$ represents the thermal noise-current due
to the resistor $R$ given by the thermodynamic dissipation-fluctuation
relationship \cite{Parisi88} 
\begin{eqnarray}
\left\langle N(t)\right\rangle &=&0  \label{eq:Eq_2} \\
\left\langle N(t)N(t^{\prime })\right\rangle & = & 2\frac{k_BT}R\delta
(t-t^{\prime })\;,  \label{eq:Eq_3}
\end{eqnarray}
with $T$ being the temperature. The symbol, $\delta (t-t^{\prime })$, is the
Dirac delta function. Current and time are usually normalized respectively
to the Josephson critical current $I_c$ and to $(\omega_0)^{-1}$, where $%
\omega_0=\sqrt{2eI_c/\hbar C}$ is the Josephson plasma frequency. With this
normalization, the coefficient of the first-order phase derivative becomes
the normalized dissipation $\alpha =\hbar \omega _0/2eRI_c$. It is also
convenient to scale the energies to the Josephson energy $E_J=I_c\hbar
/2e=I_c\Phi_0/2\pi $, where $\Phi _0=h/2e=2.07\cdot 10^{-15}Wb$ is the
flux-quantum. Thus, the set of equations (1-3) can be expressed in
normalized form as 
\begin{eqnarray}
\ddot \varphi +\alpha \dot \varphi +\sin \varphi &=&\eta +\eta _d\sin \Omega
_d\tau +n(\tau )  \label{eq:Eq_4} \\
\left\langle n(\tau )\right\rangle &=&0  \label{eq:Eq_5} \\
\left\langle n(\tau )n(\tau ^{\prime })\right\rangle &=&2\alpha \theta
\delta (\tau -\tau ^{\prime })\ \;,  \label{eq:Eq_6}
\end{eqnarray}
where $\theta =\frac{k_BT}{E_J}$ is the normalized temperature.

Figure 2 shows the results of numerical simulations of escape in a system
with $\alpha=0.00845$, $\theta =4.76\cdot 10^{-4}$, and continuous bias
sweep rate $\frac{d\eta }{d\tau }=2.1\cdot 10^{-8}$. These parameters have
been chosen in correspondence with the experimental measurements discussed
below. The resulting switching distributions (each corresponding to 10000
events), obtained for different values of the normalized drive frequency,
are shown as a function of the continuous bias, and we clearly observe that
secondary peaks in the distribution may appear, along with the primary peak
corresponding to the unperturbed ($\eta _d=0$) switching distribution. The
bias location of the secondary peak of each distribution is pointing by a
dotted line to the intersection (marked with $\bullet$) with the driving
frequency on the top plot of Figure 2. Also on the top plot, we have shown
the linear plasma resonance curve, $\Omega_p=(1-\eta^2)^{1/4}$, as a solid
line, and we observe that the numerically obtained switching distribution
peaks are closely aligned with the linear plasma resonance near the driving
frequency.

Notice that the values of the ac drive amplitude $\eta_d$ were
tuned (as indicated at the right margin of each plot) in order to achieve
the resonant energy necessary to escape the potential well. Naturally, the
lowest values of dc bias current need larger ac amplitudes to produce a
significant number of switching events. The fact that the data points on the
top graph are consistently below the predicted (linear) resonance curve can
be directly attributed to the anharmonicity of the Josephson potential for
large bias currents \cite{tobepub}.

Thermal escape properties of the Josephson junction, as described by
Eqs.~(4-6), depend strongly on the value of the normalized frequency $\Omega
_d$. We have repeated the same numerical procedure as described in Figure 2
for drive frequencies equal to sub- and superharmonics of the frequencies
leading to the direct resonances observed above. Specifically, Figure 3
reports results for driving frequencies $\Omega _d=q\Omega _p$ with $q=\frac 
15,\frac 14,\frac 13,\frac 12,1,2$ (notice that $q=1$ is the case shown in
Figure 2). The numerical results for both super- and subharmonics are shown
in Figure 3 (circles $\circ $). In this figure we see that, even when
driving at subharmonics or superharmonics, the positions of the secondary
switching peaks follow closely what can be expected from the plasma
resonance dependence on the bias current. It is worth noting that the escape
histograms often revealed multi-peaked distributions with more than two
peaks. An example of this is shown in the lowest plot of Figure 2 where we
see two peaks close to each other and another well separated on the left.
These are merely an expression of the fact that the variation in the bias
current is tuning the plasma frequency, which in turn may find some rational
resonance with the applied frequency, thereby enhancing the induced
amplitude of oscillation in the junction and increase the probability for
escape.

Complementary experiments were performed on Josephson tunnel junctions
fabricated according to classical Nb-NbAlOx-Nb procedures \cite{fabrication}%
. The samples had very good current-voltage characteristics and magnetic
field diffraction patterns. The junctions were cooled in a $^3$He
refrigerator (Oxford Instruments Heliox system), providing temperatures
down to 360mK. The microwave radiation, brought to the chip-holders by a
coax cable, was coupled capacitively to the junctions. The junction had a
maximum critical current of $143\mu A$ and a total capacitance of $6pF$ from
which we estimate a plasma frequency of $42.5GHz$. From this value of the
(unbiased) plasma frequency the classical to quantum crossover temperature 
\cite{Affleck} $T^{*}=(\hbar \omega _0/2\pi k_B)=320mK$ between classical
thermal and quantum mechanical behavior can be estimated. The sweep rate of
the continuous current $I_{dc}$ was $\dot I_{dc}=800mA/s$, and we verified
that the experiment was being conducted in adiabatic conditions \cite
{Silvestrini97}. At the temperature $T=1.6K$ the junction had a Josephson
energy $E_J=46.4\cdot 10^{-21}J$, and effective resistance $R=74\Omega $. We
have gathered data in the temperature range of $360mK-1.6K$, but we will
here show the results for $T=1.6K$. As mentioned above, simulations were
performed on the basis of these real system parameters. The evaluation of
the dissipation parameter was based on the hysteresis of the
current-voltage characteristics of the junctions\cite{Barone82,Vanduzer98};
we anticipate that the dissipation parameter used in the simulations may not
completely represent the experimental dissipation.

Figure 4 displays the experimentally obtained complementary results to
Figure 2. It is clear that also the experiments exhibit close agreement
between the bias location of the secondary switching current and the plasma
frequency resonance when the value of drive frequency is near the plasma
frequency resonance. We note that the level of the applied power is not a
simple linear function (like in Figure 2) because resonances in the coupling
system caused a slightly irregular coupling from the microwaves to the
junction. In Figure 4 each plot is obtained by recording 2000 switching
events.

Similarly to the simulations, we have also conducted escape measurements
using subharmonic microwave radiation. These data are shown in Figure 3 as
squares ($\Box $) alongside the corresponding simulation results discussed
above. In the experiments, as in the simulations, we see that the position
of the secondary peak follows closely what would be expected from the plasma
frequency dependence on the dc current bias. We emphasize that no fitting
was used in Figure 3 for neither numerical/experimental data nor the plasma
resonance curves. Data obtained at different temperatures in the range $%
(360mK-1.6K)$ showed features identical to those of Figures 2-4 for both
simulations and experiments. As was the case in the simulations, we also
frequently observed switching distributions with three peaks in the
experiments.

In conclusion, we have reported on numerical simulations and experiments on
ac-driven, thermal escape of a {\it classical} particle from a
one-dimensional potential (the Josephson washboard potential) well. We have
found that resonant coupling (harmonic or subharmonic) between the applied
microwaves and the plasma resonance frequency of the junction provides an
enhanced opportunity for escape, and we have directly observed the
signatures of such microwave-induced escape distributions in the form of
anomalous multi-peaked escape statistics. The straightforward agreement
between the classical hypothesis of anomalous distributions being directly
produced by ac-induced resonances, the results of numerical simulations of
the classical pendulum model of a Josephson junction, and actual Josephson
junction experiments indicate a consistent interpretation of ac-induced
anomalous multi-peaked switching distributions in the classical regime of
Josephson junctions.

It is noted that previous experimental work on ac-induced escape
distributions obtained at temperatures below $T^{*}$ is consistent with the
observations presented here. Those experiments have produced ac-induced
peaks in the observed switching distributions, and the relevant peaks are
located alongside the expected classical plasma resonance curve, as we have
also found here. An important observation is that the microwave-radiation
frequency necessary for populating an excited quantum level ($\hbar \omega _d
$) in a quantum oscillator coincides with the classical resonance frequency
of the corresponding classical oscillator. Thus, the switching distributions
obtained from classical and quantum mechanical oscillators may exhibit the
same microwave induced multi-peak signatures, which in the classical
interpretation is merely due to resonant nonlinear effects. The experimental
and numerical evidence herein reported, along with data existing below $%
T^{*}$ \cite{Ustinov03}, point again to the Josephson junction as an ideal
vehicle for probing relevant issues in fundamental physics \cite{Averbukh98}.

This work was supported in part by the Computational Nanoscience Group,
Motorola, Inc. We are happy to acknowledge Prof. Alan Laub for a careful
reading of the manuscript.

\begin{figure}[tbp]
\caption{Sketch of the physical phenomenon under investigation: a driven
oscillation energy $E_{ac}$ superimposed onto thermal excitations, may cause
a particle to escape a washboard potential.}
\label{fig:fig1}
\end{figure}

\begin{figure}[tbp]
\caption{Simulated switching distributions, $\rho(\eta)$, for the ac-driven
junction model obtained for increasing values of the drive frequency. The
frequency data points in the uppermost plot are relative to the position of
the secondary peak in the plots. Normalized temperature is $%
\theta=4.76\cdot10^{-4}$, bias sweep rate is $\dot{\eta}=2.1\cdot10^{-8}$,
and damping coefficient is $\alpha=0.00845$. Solid curve in uppermost graph
represents the linear plasma resonance $\Omega_p$.}
\label{fig:fig2}
\end{figure}

\begin{figure}[tbp]
\caption{The functional dependencies of the driving frequency upon the
location of the secondary peak in $\rho(\eta)$ obtained for subharmonic and
harmonic pumping. Circles ($\circ$) represent numerical results and squares (%
$\Box$) experimental data. Parameters are as given in Figure 2. Solid curves
represent $q\Omega_p$, for different $q$, as indicated.}
\label{fig:fig3}
\end{figure}

\begin{figure}[tbp]
\caption{The experimental complement to Figure 2 for a thermodynamic
temperature ($1.6K$) equivalent to the numerical normalized value.}
\label{fig:fig4}
\end{figure}

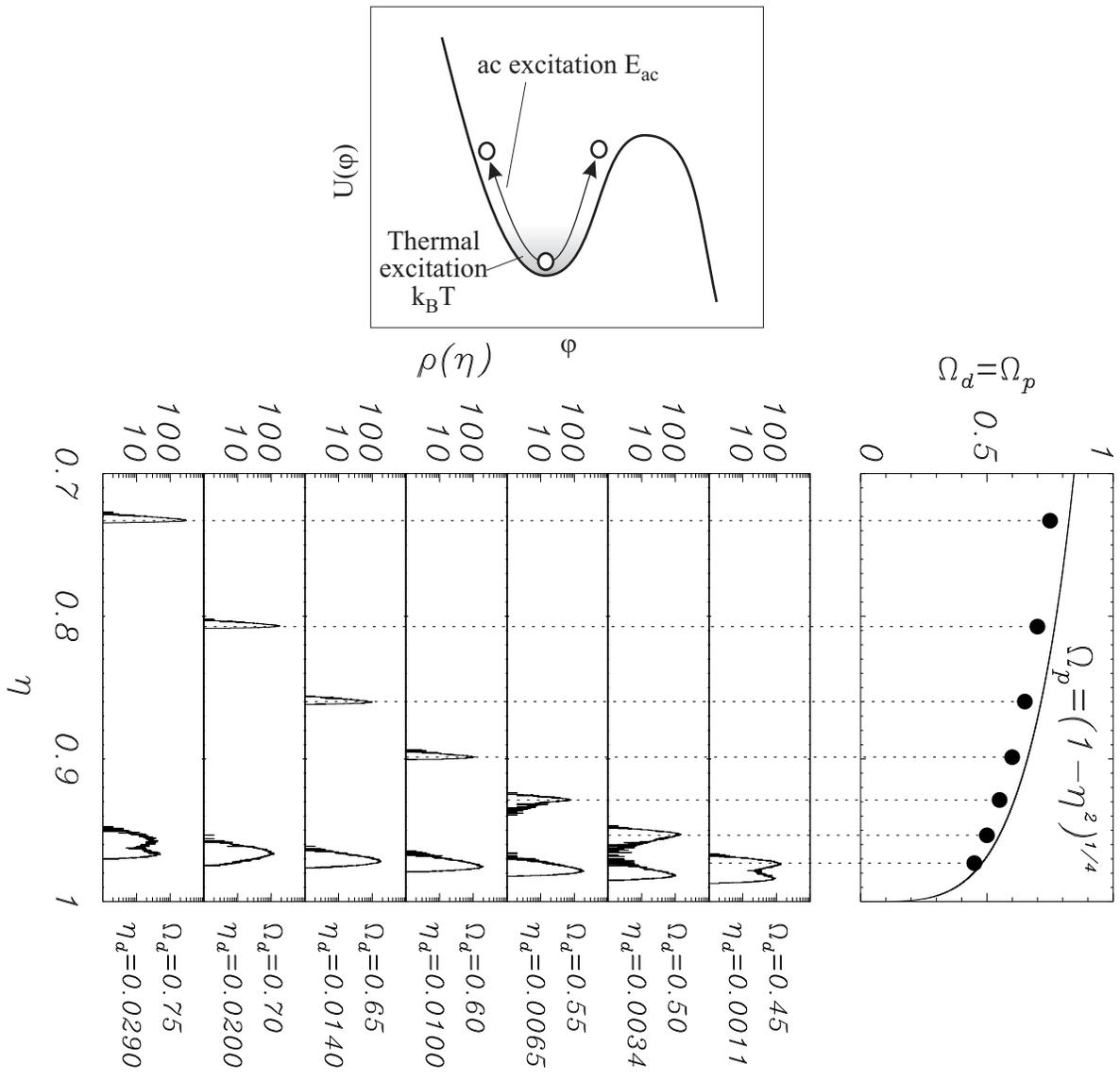
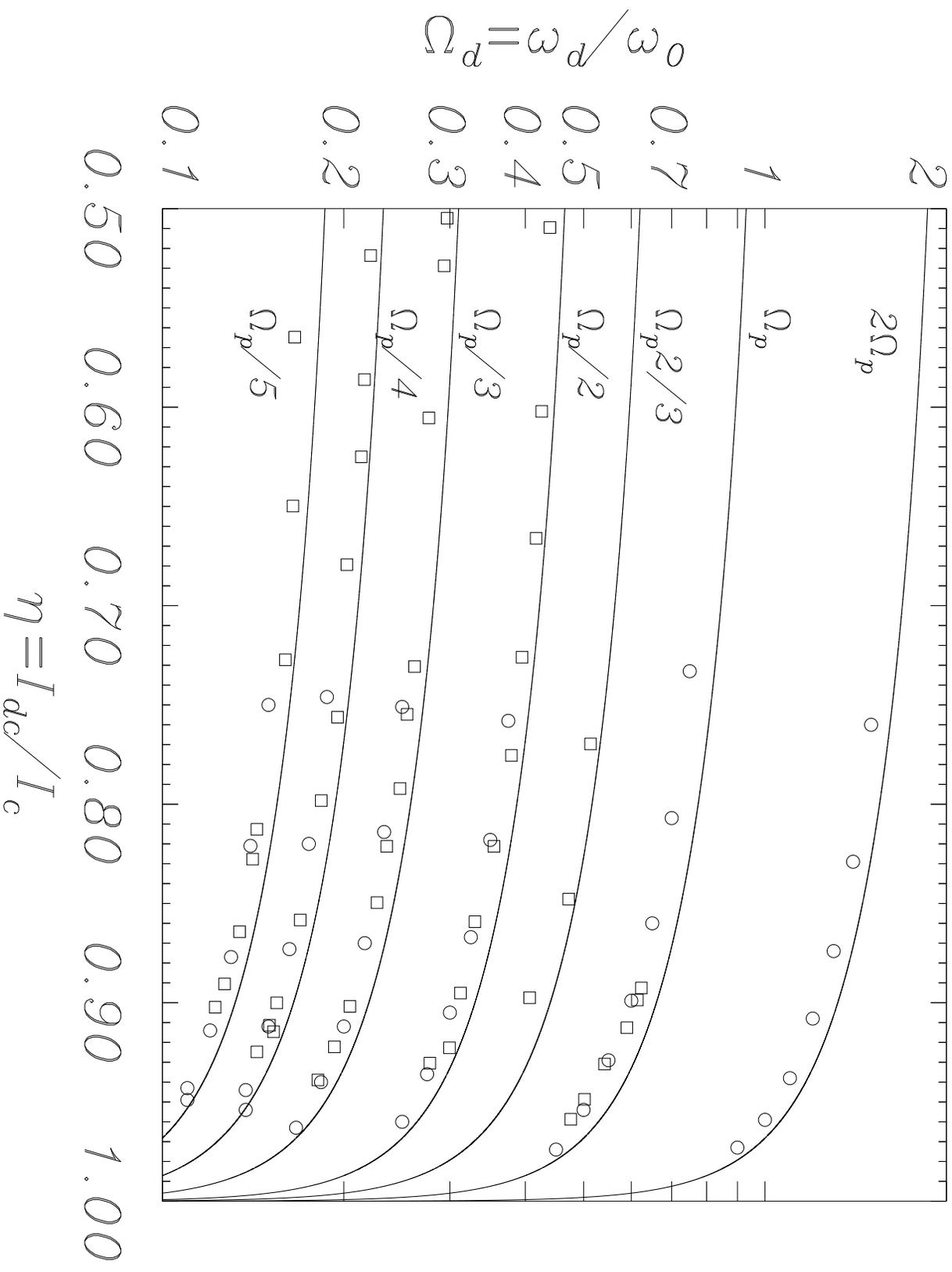
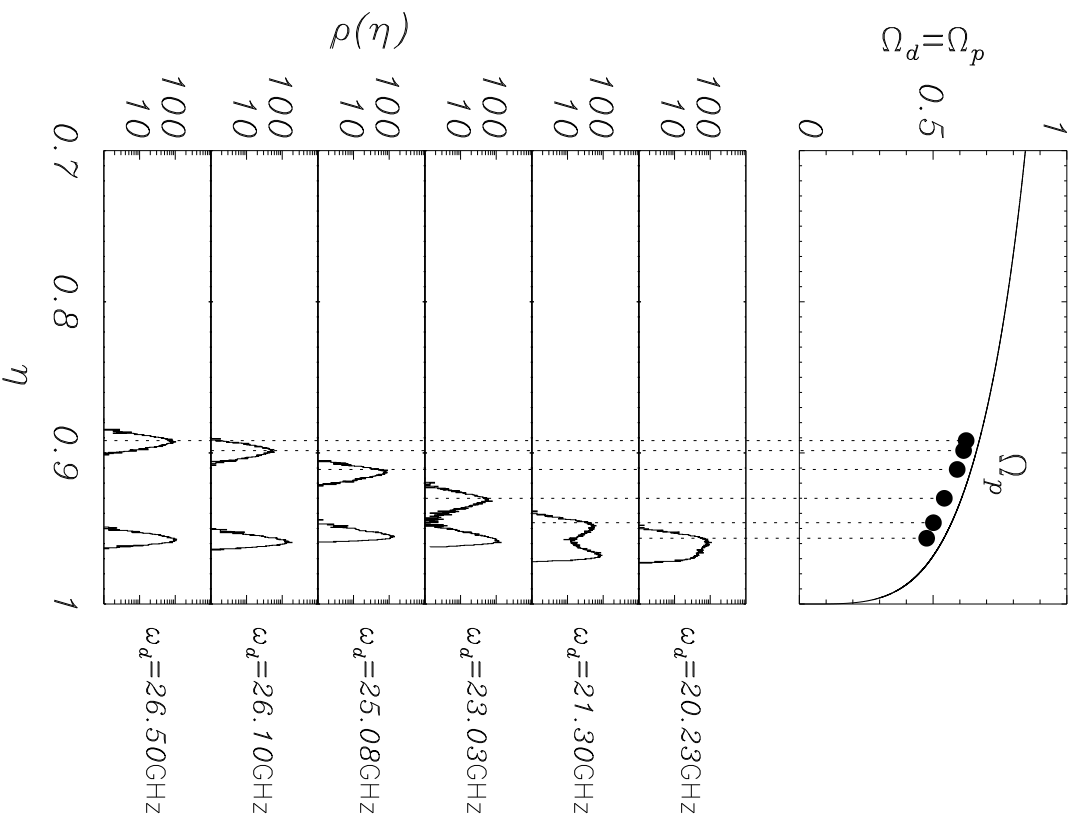
\end{document}